\documentclass[twocolumn]{aastex62}

\shorttitle{Medium-band observation of TXS 0506+056}
\shortauthors{Hwang et al.}

\begin{document}

\title{Medium-band observation of the neutrino emitting blazar, TXS 0506+056}

\author[0000-0002-3199-4234]{Sungyong Hwang}
\affil{Astronomy Program, Department of Physics \& Astronomy, Seoul National University,\\
1 Gwanak-ro, Gwanak-gu, Seoul 08826, Korea, myungshin.im@gmail.com}
\affil{SNU Astronomical Research Center, Seoul National University, 1 Gwanak-ro, Gwanak-gu, Seoul 08826, Korea}

\author[0000-0002-8537-6714]{Myungshin Im}
\affil{Astronomy Program, Department of Physics \& Astronomy, Seoul National University,\\
1 Gwanak-ro, Gwanak-gu, Seoul 08826, Korea, myungshin.im@gmail.com}
\affil{SNU Astronomical Research Center, Seoul National University, 1 Gwanak-ro, Gwanak-gu, Seoul 08826, Korea}

\author[0000-0002-0992-5742]{Yoon Chan Taak}
\affil{Astronomy Program, Department of Physics \& Astronomy, Seoul National University,\\
1 Gwanak-ro, Gwanak-gu, Seoul 08826, Korea, myungshin.im@gmail.com}
\affil{SNU Astronomical Research Center, Seoul National University, 1 Gwanak-ro, Gwanak-gu, Seoul 08826, Korea}

\author{Insu Paek}
\affil{Astronomy Program, Department of Physics \& Astronomy, Seoul National University,\\
1 Gwanak-ro, Gwanak-gu, Seoul 08826, Korea, myungshin.im@gmail.com}
\affil{SNU Astronomical Research Center, Seoul National University, 1 Gwanak-ro, Gwanak-gu, Seoul 08826, Korea}

\author{Changsu Choi}
\affil{Astronomy Program, Department of Physics \& Astronomy, Seoul National University,\\
1 Gwanak-ro, Gwanak-gu, Seoul 08826, Korea, myungshin.im@gmail.com}
\affil{SNU Astronomical Research Center, Seoul National University, 1 Gwanak-ro, Gwanak-gu, Seoul 08826, Korea}

\author[0000-0002-2188-4832]{Suhyun Shin}
\affil{Astronomy Program, Department of Physics \& Astronomy, Seoul National University,\\
1 Gwanak-ro, Gwanak-gu, Seoul 08826, Korea, myungshin.im@gmail.com}
\affil{SNU Astronomical Research Center, Seoul National University, 1 Gwanak-ro, Gwanak-gu, Seoul 08826, Korea}

\author{Sang-Yun Lee}
\affil{Astronomy Program, Department of Physics \& Astronomy, Seoul National University,\\
1 Gwanak-ro, Gwanak-gu, Seoul 08826, Korea, myungshin.im@gmail.com}

\author{Tae-Geun Ji}
\affil{School of Space Research and Institute of Natural Sciences, Kyung Hee University, \\
1732 Deogyeong-daero, Giheung-gu, Yongin-si, Gyeonggi-do 446-701, Korea}

\author[0000-0002-2548-238X]{Soojong Pak}
\affil{School of Space Research and Institute of Natural Sciences, Kyung Hee University, \\
1732 Deogyeong-daero, Giheung-gu, Yongin-si, Gyeonggi-do 446-701, Korea}

\author[0000-0003-0871-3665]{Hye-In Lee}
\affil{School of Space Research and Institute of Natural Sciences, Kyung Hee University, \\
1732 Deogyeong-daero, Giheung-gu, Yongin-si, Gyeonggi-do 446-701, Korea}

\author{Hojae Ahn}
\affil{School of Space Research and Institute of Natural Sciences, Kyung Hee University, \\
1732 Deogyeong-daero, Giheung-gu, Yongin-si, Gyeonggi-do 446-701, Korea}

\author{Jimin Han}
\affil{School of Space Research and Institute of Natural Sciences, Kyung Hee University, \\
1732 Deogyeong-daero, Giheung-gu, Yongin-si, Gyeonggi-do 446-701, Korea}

\author{Changgon Kim}
\affil{Department of Astronomy and Space Science, Kyung Hee University, \\
1732 Deogyeong-daero, Giheung-gu, Yongin-si, Gyeonggi-do 446-701, Korea}

\author{Jennifer Marshall}
\affil{Mitchell Institute for Fundamental Physics and Astronomy and Department of Physics and Astronomy, Texas A\&M University, College Station, TX 77843-4242, USA\\}

\author{Christopher M. Johns-Krull}
\affil{Department of Physics \& Astronomy, Rice University, 6100 Main Street, Houston, TX 77005, USA}

\author{Coyne A. Gibson}
\affil{McDonald Observatory, 
3640 Dark Sky Dr, Fort Davis, TX 79734, USA}

\author[0000-0001-9009-1126]{Luke Schmidt}
\affil{Mitchell Institute for Fundamental Physics and Astronomy and Department of Physics and Astronomy, Texas A\&M University, College Station, TX 77843-4242, USA\\}

\author{Travis Prochaska}
\affil{Mitchell Institute for Fundamental Physics and Astronomy and Department of Physics and Astronomy, Texas A\&M University, College Station, TX 77843-4242, USA\\}



\begin{abstract}

TXS 0506+056 is a blazar that has been recently identified as the counterpart of the neutrino event IceCube-170922A. Understanding blazar type of TXS 0506+056 is important to constrain the neutrino emission mechanism, but the blazar nature of TXS 0506+056 is still uncertain. As an attempt to understand the nature of TXS 0506+056, we report the medium-band observation results of TXS 0506+056, covering the wavelength range of 0.575 to 1.025 $\mu$m. The use of the medium-band filters allow us to examine if there were any significant changes in its spectral shapes over the course of one month and give a better constraint on the peak frequency of synchrotron radiation with quasi-simultaneous datasets. The peak frequency is found to be $10^{14.28}$ Hz, and our analysis shows that TXS 0506+056 is not an outlier from the blazar sequence. As a way to determine the blazar type, we also analyzed if TXS 0506+056 is bluer-when-brighter (BL Lac type and some flat spectrum radio quasars, FSRQs) or redder-when-brighter (found only in some FSRQs). Even though we detect no significant variability in the spectral shape larger than observational error during our medium-band observation period, the comparison with a dataset taken at 2012 shows a possible redder-when-brighter behavior of FSRQs. Our results demonstrate that medium-band observations with small to moderate-sized telescopes can be an effective way to trace the spectral evolution of transients such as TXS 0506+056.

\end{abstract}



\section{Introduction} \label{sec:intro}
 There has been a significant amount of efforts to understand the origin of high-energy cosmic neutrinos with energies $> 100$ TeV detected by the IceCube collaboration (e.g., Aartsen et al. 2013, 2014, 2019, 2020; IceCube Collaboration, 2013; Kankare et al. 2019), and their connection with high-energy astrophysical phenomena such as cosmic rays. In this regard, the recent identification of TXS 0506+056, a blazar at $z=0.3365$ (Paiano et al. 2018), as the likely site of the high-energy neutrino emission phenomenon IceCube-170922A marked the first time that enhanced electromagnetic (EM) radiation from an astronomical source is found to coincide with a high-energy cosmic neutrino event, and thus provided an exciting opportunity to understand where and how high-energy cosmic neutrinos are produced. Studies to convincingly associate TXS 0506+056 with the neutrino event is still ongoing (e.g., Lipunov et al. 2020), and so as the efforts to explain the physical mechanism for the observed neutrino and EM radiation event all together (e.g., Ansoldi et al. 2018; Padovani et al. 2018; Sahakyan. 2018; Banik \& Bhadra. 2019; Cerruti et al. 2019; Righi et al. 2019; Xue et al. 2019; Cao et al. 2020; Li et al. 2020; Morokuma et al. 2020; Petropoulou et al. 2020; Zhang et al. 2020). One of key ingredients for such efforts to succeed is to understand the physical properties of the neutrino emission site, namely the blazar TXS 0506+056.
 
 The leading candidate of high-energy neutrino site has been blazars. Therefore, it is not very surprising that IceCube-170922A was found to be associated with one. Blazars are active galactic nuclei (AGNs) with their relativistic jets pointing toward us. The production of high-energy neutrinos is thought to be a result of the interaction of protons and nuclei accelerated in the jet with the surrounding medium that triggers chain reactions and produce high-energy neutrinos and cosmic rays. The same environment provides relativistically traveling leptons that give rise to the prominent two components in the spectral energy distribution (SED) of blazars, the low-energy peaked component (the infrared to X-ray) through synchrotron radiation of high-speed electrons, and the high-energy peaked component (GeV to TeV gamma-rays) through inverse Compton scattering of low-energy photons by electrons.
 
 However, there is a wide variety of blazars and different types of blazars often represent very different physical environments, making it important to pin down the dominant type of neutrino emitting blazars for constraining the high-energy neutrino emission mechanism. Blazars can be broadly divided into two types, BL Lac objects and Flat Spectrum Radio Quasars (FSRQs). Phenomenologically speaking, BL Lacs are blazars with featureless optical spectra, while FSRQs have strong, quasar-like emission lines in their optical spectra (Padovani et al. 2017). Physically speaking, FSRQs are intrinsically high-luminosity AGNs with their jets pointing toward us, and BL Lacs are intrinsically low-luminosity AGNs such as radio galaxies with beamed jets very closely aligned to us. Therefore, in terms of energetics, FSRQs are more likely to be the site of high-energy neutrino emission (e.g., Murase et al. 2014; Dermer et al. 2014), although there are proposed mechanisms that can produce high-energy neutrinos from BL Lacs (Tavecchio et al. 2014; Righi et al. 2017; Ansoldi et al. 2018) or even from radio galaxies (e.g., Hooper 2016).
 
 Interestingly, the optical spectrum of TXS 0506+056 shows no notable features and a smooth continuum indicative of the emission being dominated by synchrotron radiation (Halpern et al. 2003; Paiano et al. 2018). Therefore, TXS 0506+056 is a BL Lac object from the spectral classification point of view. However, Padovani et al. (2019, hereafter, P19) recently suggested that TXS 0506+056 is a masquerading BL Lac with properties in line with FSRQs based on several accounts: (i) its radio and [OIII] luminosities are consistent with those of jetted luminous quasars rather than radio galaxies; (ii) it has the emission line ratios of Seyfert 2 galaxies; and (iii) the Eddington ratio is $> 0.01$, which is too high for BL Lacs and low luminosity AGNs. Furthermore, they showed that TXS 0506+056 is an outlier of the blazar sequence, a sequence defined in the 2-D plane of the peak frequency of the synchrotron radiation of blazar SED, $\nu^{S}_{\rm{peak}}$, versus the gamma-ray luminosity (Ghisellini et al. 1998; Fossati et al. 1998; Giommi et al. 2012; Padovani et al. 2012). Therefore, it is necessary to examine in detail the observational characteristics of TXS 0506+056 to see how it fits into the blazar types.
 
 One way to better constrain the blazar nature of TXS 0506+056 is to study its SED in the optical-infrared (IR), and spectral variability as a function of time. The $\nu^{S}_{\rm{peak}}$ has been found to be $\lesssim 10^{14}$ Hz for FSRQs, and $\gtrsim 10^{14}$ Hz for masquerading BL Lacs. Previous studies looked into this issue already, finding $\nu^{S}_{\rm{peak}} \sim 10^{14.5}$ Hz using available archival data from various epochs and sparsely sampled photometric information at around the IceCube-170922A event epoch (e.g., P19). However, re-examination of  the $\nu^{S}_{\rm{peak}}$ value using quasi-simultaneous, densely sampled (in wavelength) datasets are desirable to better constrain $\nu^{S}_{\rm{peak}}$ since TXS 0506+056 is known to show a rapid variability and the existing data are sometimes too sparse in wavelength coverage. Optical spectral variability of blazars have been investigated extensively in the past, and it has been found that the optical spectra of BL Lacs are ``bluer-when-brighter'' (Ikejiri et al. 2011; Bonning et al. 2012; Gaur et al. 2012; Wierzcholska et al. 2015), while a significant portion of FSRQs ($\sim$ tens of \%) have the variability of ``redder-when-brighter'' (Gu et al. 2006; Hu et al. 2006; Rani et al. 2010; Meng et al. 2018). Although the fraction of ``redder-when-brighter'' FSRQs may be lower and some FSRQs show ``bluer-when-brighter'' or no variable behavior, virtually no BL Lacs are found to be ``redder-when-brighter''. Therefore, the behavior of the optical spectral variability can be another indicator to tell if TXS 0506+056 is a BL Lac or a FSRQ.
 
 In this respect, low resolution spectroscopy with narrow or medium-band (MB) filters can be a powerful way to gain insight on the characteristics of the blazar. Although moderate- to high-resolution spectra can give us rich information about the object in study, a wider spectral width sampled by each MB filter increases signal-to-noise (S/N) per spectral element in comparison to conventional moderate- to high-resolution spectroscopy. At the same time, MB filter widths are many times finer than broad-band filters, so that shapes of SEDs can be better traced with MB filters than broad-band filters. With these advantages, small ($\lesssim 1$ m) and mid-sized telescopes ($\sim 2$ m) can be utilized for MB-based low resolution spectroscopy, even for faint objects for which spectroscopy has been considered only possible with large telescopes. Since observing time is more readily available for small and mid-sized telescopes than large telescopes, time-intensive spectroscopy of many targets, such as long-term monitoring observations, can be done with MB filters on a small telescope. Several examples that show the power of the MB observations are the identification of faint, high-redshift quasars (Jeon et al. 2017; Kim et al. 2018a, 2019b; Shin et al. 2020) and the AGN reverberation mapping (Kim et al. 2019a).
 
 To better understand the nature of the optical variability associated with IceCube-170922A, we conducted a series of MB filter observations using SED camera for Quasars in Early Universe (SQUEAN, Kim et al. 2016) and the Wide-field Integral-Field Unit (IFU) Telescope (WIT) from 2017 October 14 (UT), or about 23 days after the IceCube-170922A event, until 2017 November 6. Here, WIT is a 0.25 m wide-field imaging telescope, equipped with multiple MB filters covering a field view of 2.34 $\times$ 2.34 deg$^2$. Since it can produce MB-based low-resolution spectra at each pixel in the entire field of view, we name the telescope as the Wide-field ``IFU'' Telescope.
 
 This paper presents results from our mostly MB-based observations of TXS 0506+056, focusing on demonstration of the effectiveness of MB-based data for the study of the SED temporal variation and $\nu^{S}_{\rm{peak}}$ of the blazar emission. We introduce WIT in Section 2. The observation and data reduction are introduced in Section 3. The variability of TXS 0506+056 during our monitoring period is given in Section 4, along with the time-series SEDs of the blazar. In Section 5, we derive $\nu^{S}_{\rm{peak}}$ and discuss the spectral variability of this object based on our observations. Finally, we present our conclusion in Section 6.

\section{Wide-field Integral-Field-Unit Telescope (WIT)}
\label{sec:wit}
 WIT is a 0.25 m, f/3.6 Takahashi CCA-250 telescope equipped with a 4k $\times$ 4k CCD camera and a series of MB filters. The telescope is a modified Cassegrain telescope with a focal reducer. The CCD camera is the Finger Lake Instrumentation (FLI)'s ML16803 model, which has 4096 $\times$ 4096 pixels with a pixel size of 9 $\mu$m. Therefore, the pixel scale is $2 \farcs 06$, and the camera covers a field of view of 2.34 $\times$ 2.34 deg$^2$.
 
   \begin{figure}[h!]
 \epsscale{0.6}
\plotone{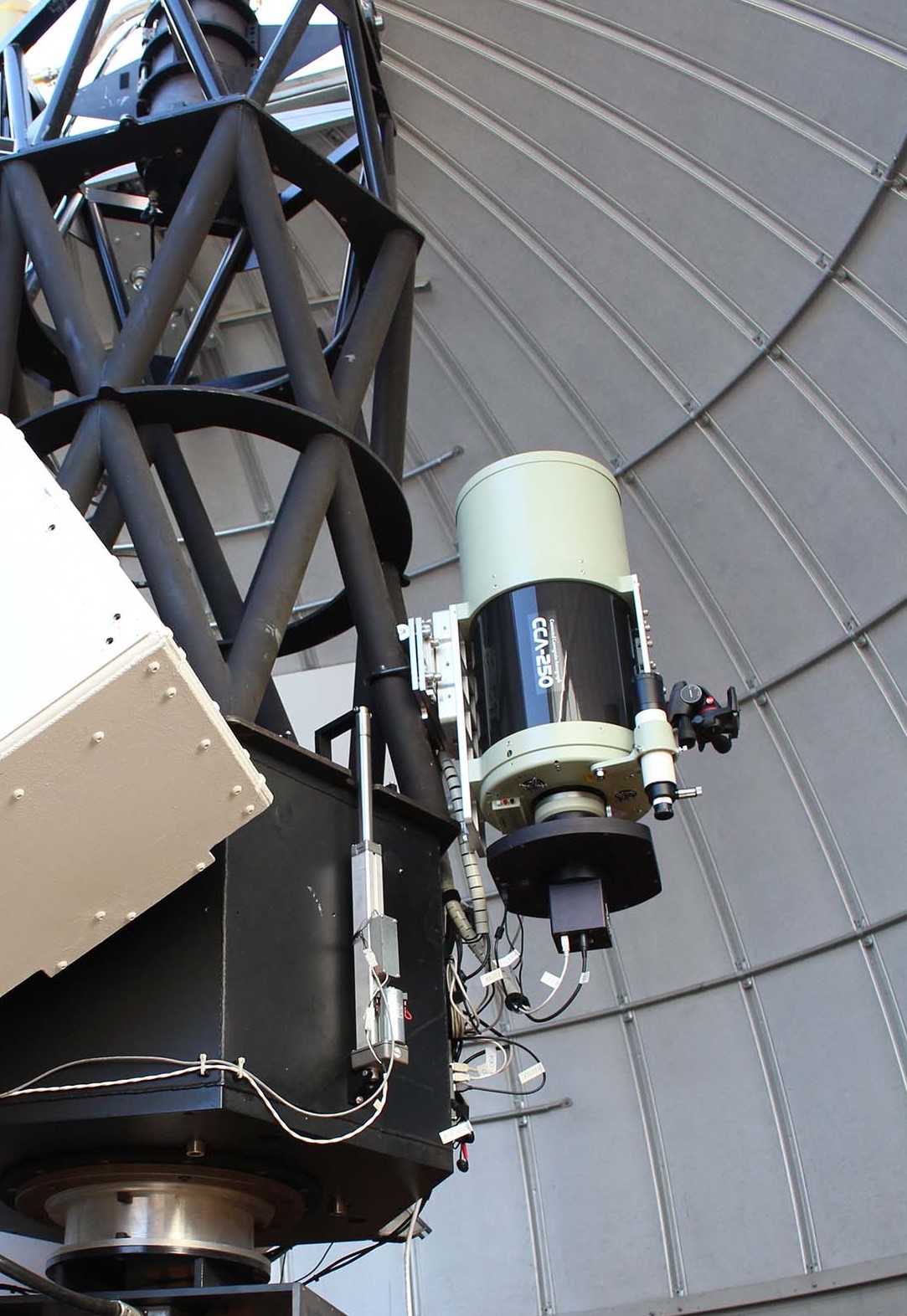}
\caption{Wide-field IFU Telescope (middle), attached on the 0.8 m telescope (left) of the McDonald Observatory. \label{fig:fig1}}
\end{figure}

 The telescope was mounted on the 0.8 m telescope of the McDonald Observatory on 2017 February 14 (Figure 1). The system includes a filter wheel that can house eight 50 mm square filters, and a suite of softwares that allows remote operation of the telescope. Currently, five medium-band filters, $m575, m625, m675, m725$, and $m775$ are installed, where the alphabet $m$ in the filter name stands for ``medium-band'', and the number represents the central wavelength of the filter in nm. The filter widths are 50 nm each. The combined transmission curves of the filters and the quantum efficiency of the camera are shown in Figure 2.  The effective wavelength and the full-width at half-maximum (FWHM) of the combined filter throughput of each filter of WIT are computed following Choi \& Im (2017), and they are presented in Table 1.
 
 \begin{deluxetable}{c|c|c}
\tablecaption{Effective wavelength and FWHM of MB on WIT\label{tab:tab1}}
\tablehead{
\colhead{Filter} & \colhead{$\lambda_{eff}$} & \colhead{FWHM}
}
\colnumbers
\startdata
$m575$ & 573.7 & 48.4\\
$m625$ & 623.6 & 48.7\\
$m675$ & 674.6 & 48.6\\
$m725$ & 724.8 & 48.8\\
$m775$ & 774.2 & 49.2\\
\enddata
\tablecomments{
(1) - Filter name;
(2) - Effective wavelength of the filter (nm);
(3) - Full-width at half-maximum (nm)}
\end{deluxetable}
 
 \begin{figure}[h!]
 \epsscale{2.5}
  \plottwo{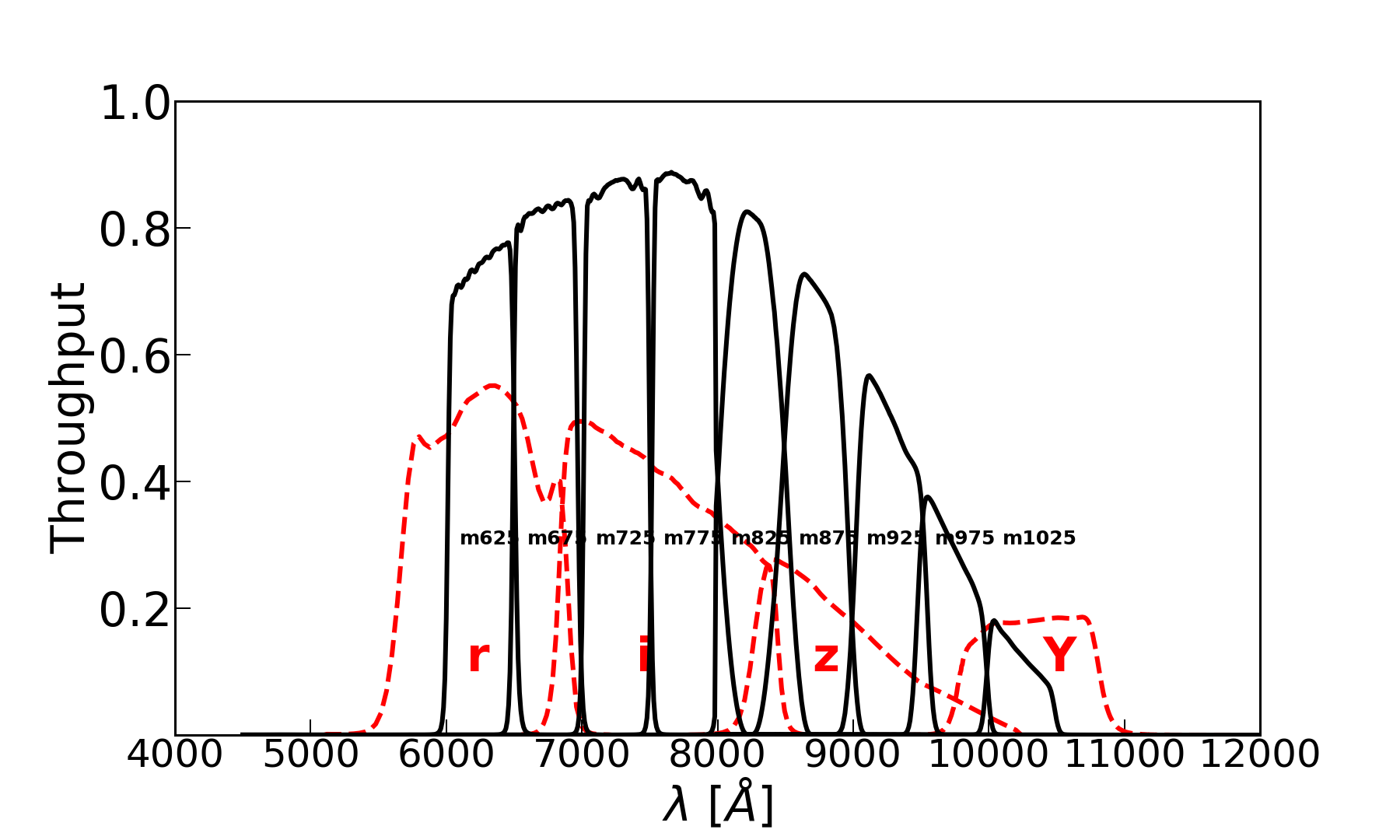}{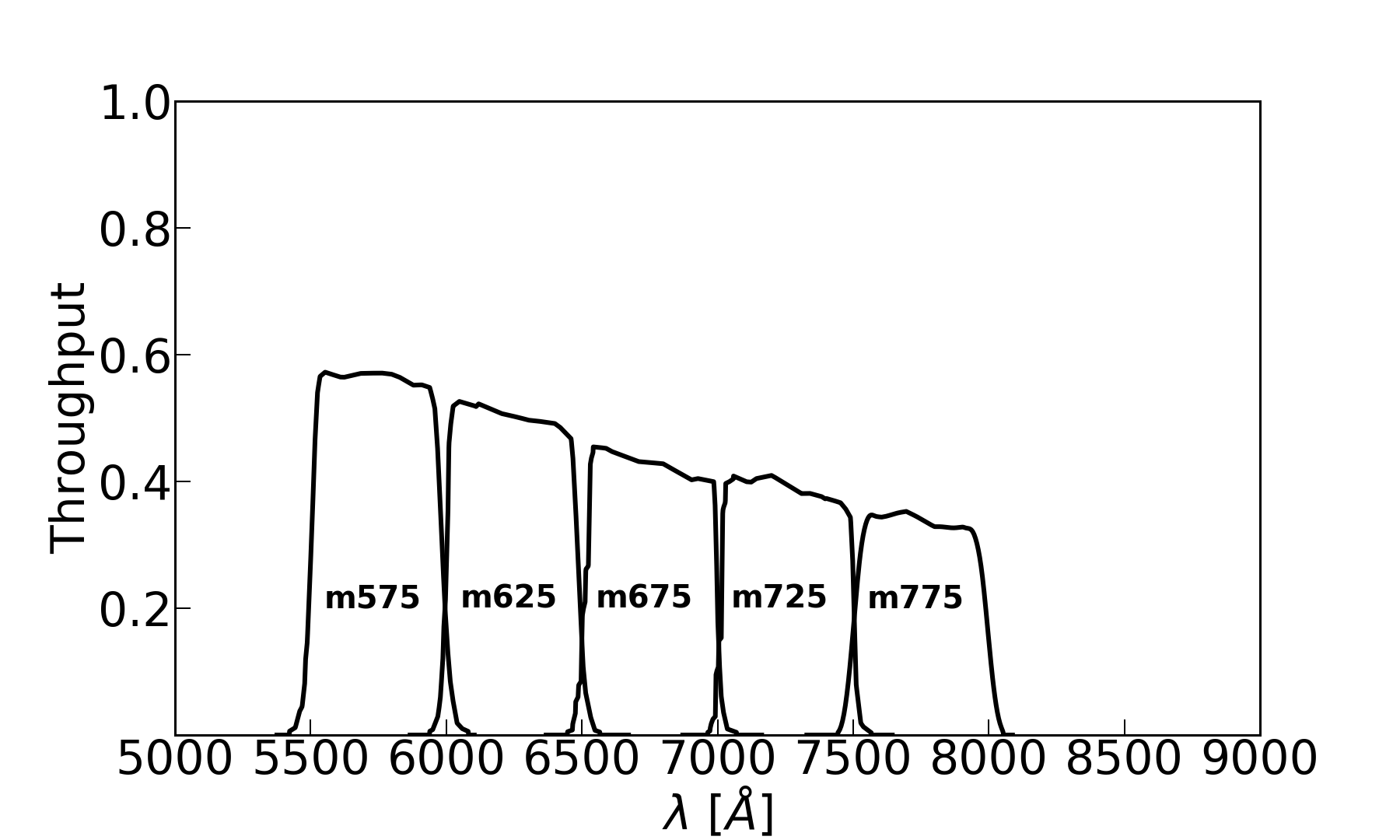}
\caption{The throughput curves for each filter of SQUEAN (Top) and WIT (Bottom). The black lines represent the throughput of MBs and the red, dashed lines represent the throughput of broad-bands. The names of filters are shown on the throughput curve.\label{fig:fig2}}
\end{figure}

 In addition to these MB filters, WIT can be equipped with the standard Johnson $B, V, R, I$ filters and holographic diffraction gratings for slitless spectroscopy. WIT has been operational since 2017 February 16, and mostly operated remotely from Korea, for monitoring observation of nearby galaxies to discover supernovae (Im et al. 2019), and to trace the spectral variation of AGNs in the field around each nearby galaxy.
 \section{Observations and data reduction} 
 \label{sec:obsdata}
\subsection{Observation}

 We observed TXS 0506+056 with the Otto Struve 2.1 m telescope and WIT at the McDonald Observatory. The 2.1 m Otto Struve telescope is equipped with SQUEAN. SQUEAN has 10 MB and 10 broad-band filters, allowing us to trace SED shapes of targets with a moderate investment of observational resources (see, e.g. Kim et al. 2019b).

 The observations were conducted at 2017 October 14 through 2017 November 6 (UT) for a total of 10 nights. We used 9 MBs and 4 broad-bands of SQUEAN, and 5 MBs for the WIT observation. The exposure time per frame was set at 150 seconds for the WIT observations. For the case of SQUEAN, the exposure times were set at 10 secs to 60 secs depending on the weather condition and the passband. The observation logs are given in Tables 5 and 6 in Appendix A.
 
\begin{figure}[ht!]
\epsscale{1.3}
\plotone{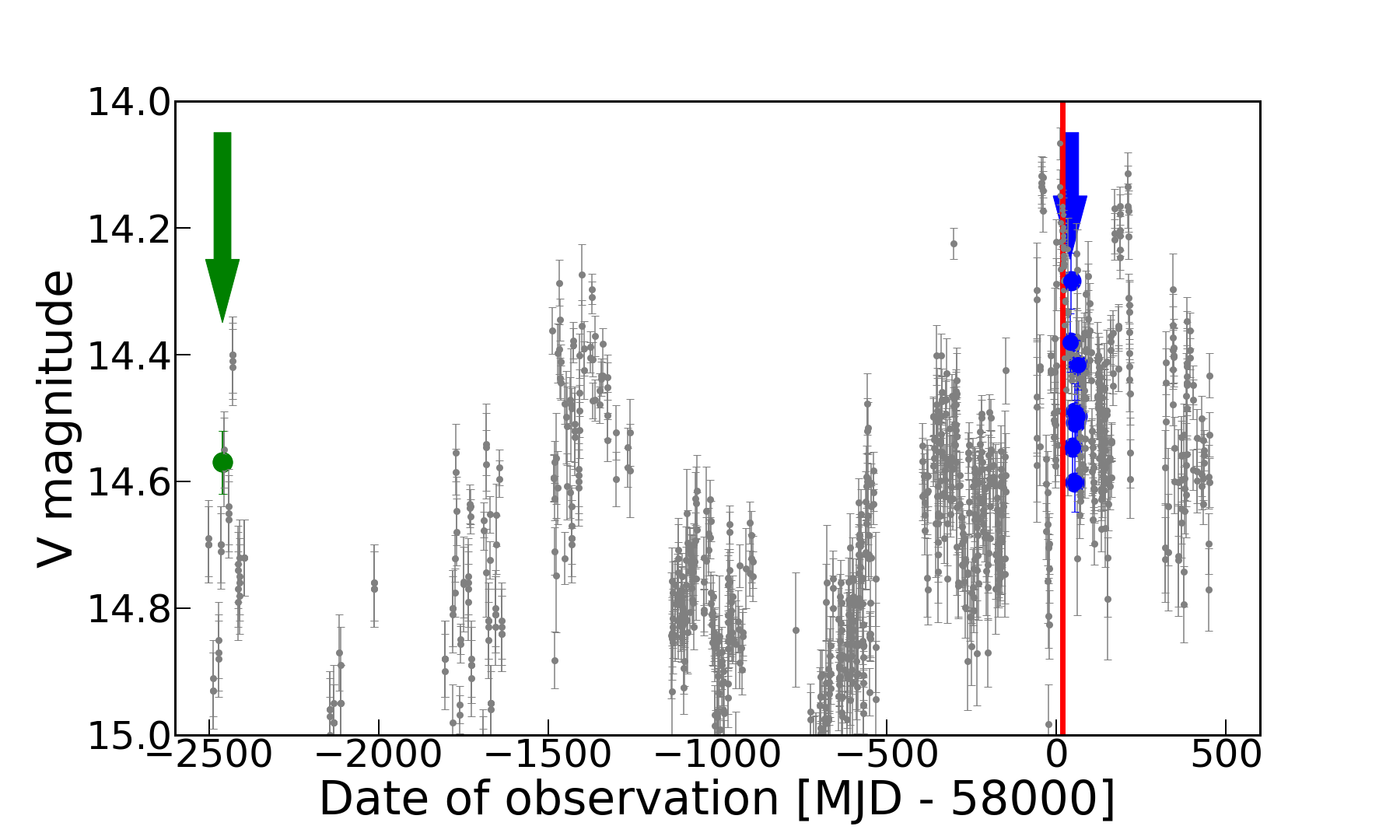}
\caption{The $V$-band light curve of TXS 0506+056 from CRTS and ASAS-SN surveys. Only the points that have magnitude error less than 0.1 are shown.  The $V$-band magnitude of the object increased about 0.5 mag from 50 days before the neutrino event. The time of the neutrino detection is shown with a red vertical line. The epoch of the Rau et al. (2012) data (inactive phase) is indicated with the green arrow and the corresponding $V$-band magnitude is marked with a green circle, while our observation period is indicated with the blue arrow and the corresponding $m575$-band magnitudes are overplotted with blue circles. \label{fig:fig9}}
\end{figure}
 
 Figure 3 indicates $m575$ data from our observations, overplotted on the Catalina Real-time Transient Survey (CRTS, Drake et al. 2009) and the All-Sky Automated Survey of Supernovae (ASAS-SN, Shappee et al. 2014; Jayasinghe et al. 2019) $V$-band light curve. Epochs of our observations correspond to the time shortly after the most active phase near the IceCube-170922A event, but still during the active phase.
 
\subsection{Data reduction and calibration} 
 We reduced the data in a standard manner, for the bias, dark, and flat-field corrections. For flat-field correction, we used both sky flat-fields and dome flat-fields. The IRAF (Tody 1986) and custom IDL codes were used for these reductions.
 
\begin{figure*}[ht!]
\includegraphics[scale=0.315]{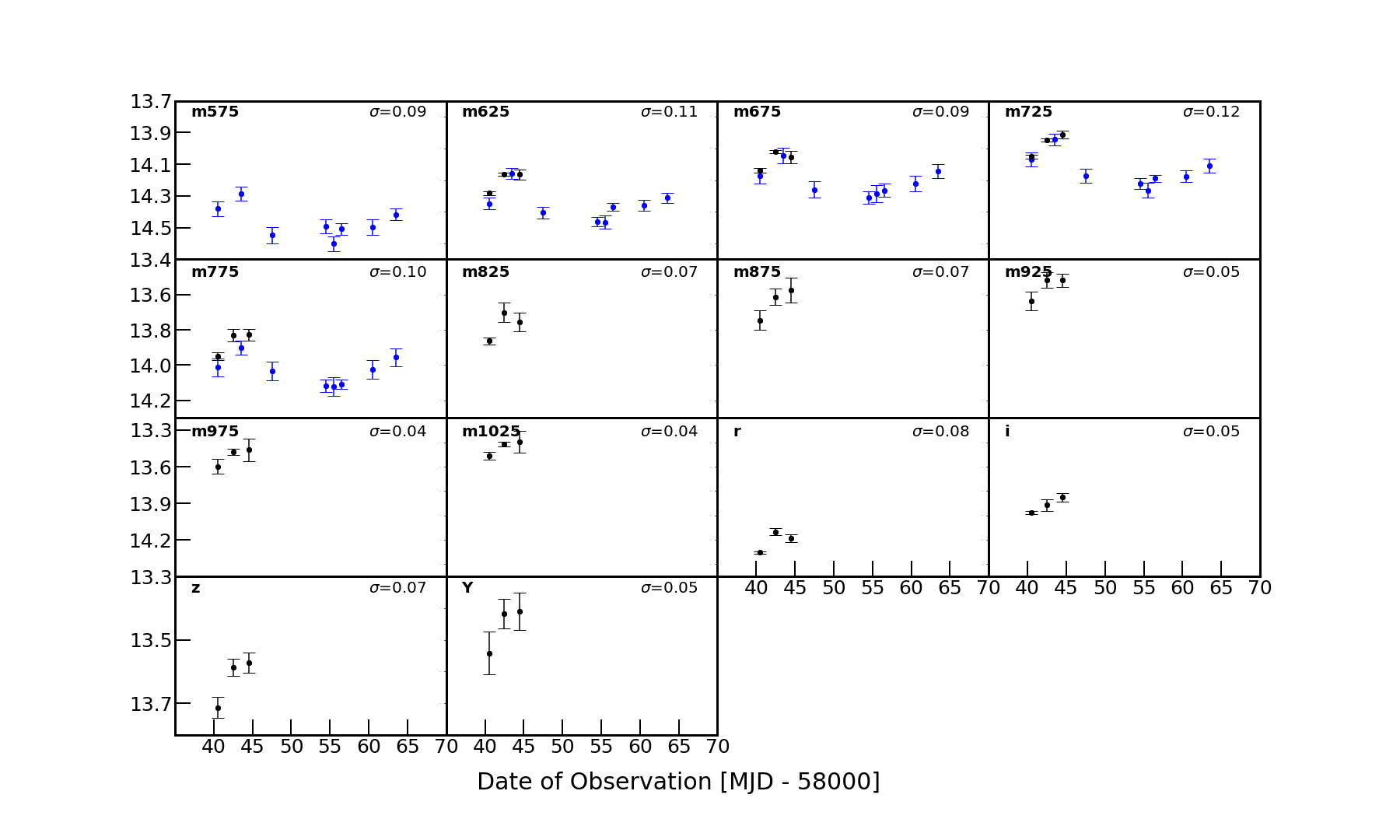}
\caption{Light curves of TXS 0506+056 using different filters. The blue points are the WIT data and the black points are those from SQUEAN. It shows the mild variabilities in the filters $m575$ through $m775$. The filter name is shown in the upper left corner and the excess variability is shown in the upper right corner in each plot.\label{fig:fig3}}
\end{figure*}
 
 Photometric calibration was done using objects near the target in the image, in a manner similar to the method described in Jeon et al. (2016) and Choi \& Im (2017). In this procedure, we first identified point sources in the field to be used as photometry reference stars by running the object detection software SExtractor (Bertin \& Arnouts 1996) and matched them to $B, V, g, r, i$ photometry from the AAVSO Photometric All Sky Survey (APASS) catalog (Henden et al. 2016). For the case of WIT, we selected 10 to 40 photometry reference stars within 36$\arcmin$ from the target, where the point sources are chosen to be those with the probability of being point source greater than 0.98, which is calculated by SExtractor. For the 2.1 m telescope data, we selected about 10 stars in the SQUEAN field of view. For both broad-band and MB data, we fitted spectral templates of 175 stars (Gunn \& Stryker, 1983; Strecker et al., 1979) to the multi-band APASS magnitudes of the reference stars, and derived the best-fit template. After identifying the best-fit template, we used the template to synthetically calculate the magnitude of each band, $m_{\rm AB}$, of the reference stars using the following equation:
 \begin{equation}
m_{\rm{AB}}= -2.5 \, \rm{log}( \frac{\textit{$\int$} \textit{d}\textit{$\lambda \, \lambda$} \, \textit{F}_{\lambda} \, \textit{R}_{\lambda}}{\textit{c} \, \textit{$\int$} \textit{d}\textit{$\lambda$} \, \textit{R}_{\lambda} \, \textit{/} \, \textit{$\lambda$}})-48.6 ,
\end{equation}
 where $R_{\lambda}$ is the throughput function of each filter including the detector quantum efficiency, and $F_{\lambda}$ is the specific flux of the best-fit stellar template. Using these magnitudes of the reference stars, we derived the zero point of each image in each filter. The error is taken as the standard deviation of the zero points from all the reference stars. To derive the magnitude of TXS 0506+056, we performed aperture photometry using SExtractor. The aperture size was chosen as 1.5 times of the seeing value to obtain the maximal signal-to-noise (S/N). The magnitudes of the reference stars were obtained in the same way, therefore, the aperture magnitude zero points include aperture correction. The photometric error of each magnitude is computed as root mean square (rms) of the zero point error and the photometric error from SExtractor.
 Note that the magnitudes in the tables are corrected for the Galactic extinction, by adopting E(B-V) = 0.094 from E(B-V) map of Schlafly $\&$ Finkbeiner (2011) with the Cardelli et al. (1989) law assuming $\it R_{V}$ = 3.1 and the dust extinction function of Fitzpatrick (1999).

\section{results} \label{sec:data}
\subsection{Variability of TXS 0506+056}
 Figure 4 shows the MB and broad-band light curves of TXS 0506+056. At a glance, TXS 0506+056 seems to be somewhat variable during the explored period, showing a decrease in flux, and then increase in flux again. To quantify the significance of variability, we performed two analyses. 
 
 First, we performed a $\chi^{2}$ test to see how variable the light curves are (de Diego 2010; Villforth et al. 2010; Kim et al. 2018b). The $\chi^{2}$ value of each light curve is
\begin{equation}
\chi^2 =
 \sum_{i=1}^{N}\frac{(m_{i}-\mu_{mag})^2}{\sigma_{mag,i}^{2}} , 
\end{equation}
 where $m_{i}$ is the magnitude at the epoch $i$, $\mu_{mag}$ is the mean value of the magnitudes, and $\sigma_{mag,i}$ is the magnitude error at the epoch $i$.
 The null hypothesis is that the object in the filter is not variable. We can calculate the confidence level of rejection, $C_{reject}$ as
\begin{equation}
C_{reject} =
 1 - \int_{\chi^2}^{\infty}P(\chi^2 ,dof) \, d\chi^2,
\end{equation} 
 where $P(\chi^2 ,dof)$ is the $\chi^{2}$ distribution fuction of the degree of freedom $dof$. The results are presented in Table 2, and the $C_{reject}$ values are about 1 for all the filters except for the filters with only 3 epochs and the mean photometric error $\langle \sigma_{mag} \rangle > 0.045$ Therefore, the $\chi^2$ result shows that the object is variable. 
 Second, in order to confirm the variability of the object, C-test value ($C_{test}$) was calculated as (e.g., Kim et al. 2018b)
\begin{equation}
C_{test} = \frac{\sigma_{\rm{LC}}}{\langle\sigma_{mag}\rangle},
\end{equation}
where $\langle\sigma_{mag}\rangle$ is the average photometric error over all the epochs, and $\sigma_{\rm{LC}}$ is the standard deviation of the light curve magnitudes. If $C_{test}$ is greater than 1.96 (2.56), the object is variable with 95\% (99\%) confidence level. We find that the $C_{test}$ values are above 1.96 and some exceed 3, except for those for the light curves of the filters at $\geq 875$ nm with only 3 epochs and $\langle\sigma_{mag}\rangle > 0.05$ mag (Table 2). The $C_{test}$ value for the $m975$ filter light curve is the lowest value of 1.25, and even for this case, we can say that the confidence level is 79\% in favor of variability.
 With the results from the $\chi^2$ test and C-test, we conclude that the blazar has been variable during our observation.
  To quantify how much the object was variable, we derived the excess variability $\sigma_{ex}$ defined as (Kim et al. 2018b, 2019a),
\begin{equation}
\sigma_{ex} =
 \sqrt{\sigma_{\rm{LC}}^{2} - \langle \sigma_{mag} \rangle^{2}} \, .
\end{equation}
The excess variabilities are found to be at 0.04 to 0.12 mag, and they are positive for all the MB filters. In particular, for the filters with more than 8 epochs of data ($m575$ through $m775$), suggesting inter-day variability of TXS 0506+056 at the level of 0.1 mag level during the monitored period.

\begin{deluxetable}{c|c|c|c|c|c|c|c}
\tablecaption{Variability Test Results\label{tab:tab2}}
\tablehead{
\colhead{Filter} & \colhead{$\chi^2$} & \colhead{$\nu$} & \colhead{$\sigma_{LC}$} & \colhead{$\langle\sigma_{mag}\rangle$} & \colhead{$C_{reject}$} & \colhead{$C_{test}$} & \colhead{$\sigma_{ex}$}
}
\colnumbers
\startdata
$m575$ & 34.8 & 8 & 0.101 & 0.045 & 1.00 & 2.24 & 0.09 \\
$m625$ & 394.8 & 11 & 0.114 & 0.029 & 1.00 & 3.93 & 0.11\\
$m675$ & 249.4 & 11 & 0.103 & 0.040 & 1.00 & 2.58 & 0.09\\
$m725$ & 392.9 & 11 & 0.122 & 0.032 & 1.00 & 3.81 & 0.12\\
$m775$ & 94.1 & 11 & 0.107 & 0.041 & 1.00 & 2.61 & 0.10\\
$m825$ & 23.3 & 3 & 0.083 & 0.042 & 1.00 & 1.98 & 0.07\\
$m875$ & 4.7 & 3 & 0.090 & 0.057 & 0.80 & 1.58 & 0.07\\
$m925$ & 4.2 & 3 & 0.069 & 0.045 & 0.76 & 1.53 & 0.05\\
$m975$ & 4.1 & 3 & 0.075 & 0.060 & 0.75 & 1.25 & 0.05\\
$m1025$ & 6.5 & 3 & 0.061 & 0.048 & 0.91 & 1.27 & 0.04\\
$r$ & 129.2 & 3 & 0.087 & 0.023 & 1.00 & 3.78 & 0.08\\
$i$ & 28.1 & 3 & 0.062 & 0.031 & 1.00 & 2.00 & 0.05\\
$z$ & 11.8 & 3 & 0.078 & 0.031 & 1.00 & 2.52 & 0.07\\
$Y$ & 2.9 & 3 & 0.074 & 0.058 & 0.59 & 1.28 & 0.05\\
\enddata
\tablecomments{
(1) - Filter name;
(2) - Chi-squared value;
(3) - Degree of freedom;
(4) - Standard deviation of the light curve (mag);
(5) - Mean photometric error of the light curve (mag);
(6) - Confidence of rejection;
(7) - C-statistic;
(8) - Excess variability (mag);
}
\end{deluxetable}

\begin{figure*}[ht!]
\includegraphics[scale=0.32]{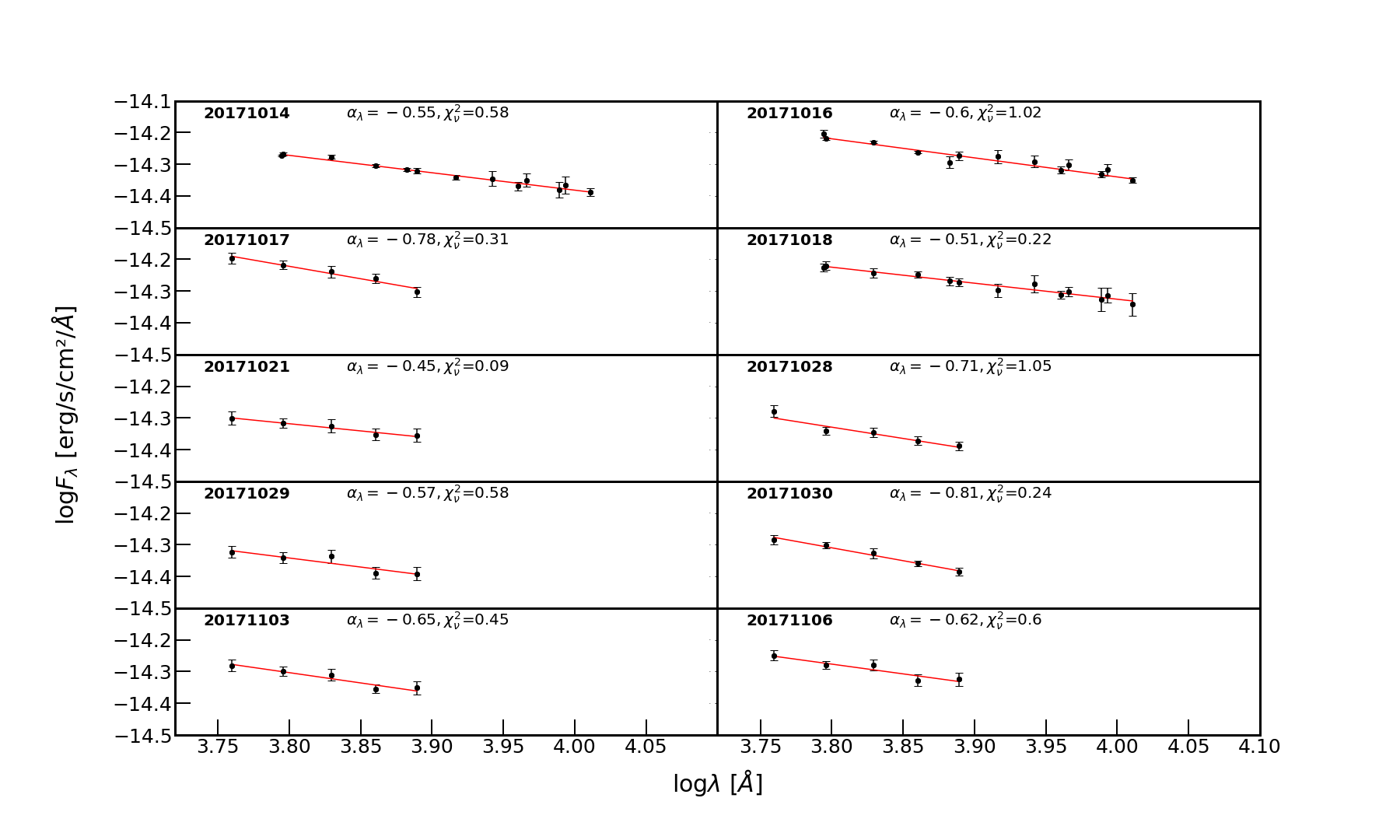}
\caption{The optical SEDs at each epoch and the fitting results. The best-fit spectral slope $\alpha_{\lambda}$ and $\chi_{\nu}^2$ values are shown in the upper right corner of each plot. UT date of the observation is shown as well.\label{fig:fig4}}
\end{figure*}
 
\subsection{SED temporal variation}

 \begin{deluxetable}{c|c|c|c}
\tablecaption{Power-law fitted parameters\label{tab:tab3}}
\tablehead{
\colhead{Date} & \colhead{$\alpha_{\lambda}$} & \colhead{$\sigma_{\alpha_{\lambda}}$} & \colhead{$\chi_{\nu}^2$}
}
\colnumbers
\startdata
2017-10-14 & -0.55 & 0.03  & 0.58\\
2017-10-16 & -0.60 & 0.03 & 1.02\\
2017-10-17 & -0.78 & 0.16 & 0.31\\
2017-10-18 & -0.51 & 0.07 & 0.22\\
2017-10-21 & -0.45 & 0.19 & 0.09\\
2017-10-28 & -0.71 & 0.15 & 1.05\\
2017-10-29 & -0.57 & 0.19 & 0.58\\
2017-10-30 & -0.81 & 0.12 & 0.24\\
2017-11-03 & -0.65 & 0.18 & 0.45\\
2017-11-06 & -0.62 & 0.17 & 0.60\\
\enddata
\tablecomments{
(1) - Date of observation (UT);
(2) - Power-law slope;
(3) - Error of the power-law slope;
(4) - Reduced chi-squared value}
\end{deluxetable}

 The time-series MB data allow us to examine in detail how the SED of TXS 0506+056 evolves with time. To quantify the variability of the SED shape, we modeled the specific flux of TXS 0506+056 with a power-law shape of $F_{\lambda} \sim
 \lambda^{\alpha_{\lambda}}$. We did so because typical blazars show a power-law shape of SEDs indicative of the synchrotron radiation (e.g., Zacharias \& Schlickeiser 2010). In particular, TXS 0506+056 has been reported as having a featureless power-law continuum in optical with the $\alpha_{\lambda}$ value at around $\alpha_{\lambda} \sim -1$ to 0 (Rau et al. 2012; IceCube Collaboration et al. 2018; Paiano et al. 2018). We fitted the observed SEDs at each epoch with this power law model,
 \begin{equation}
\rm{log} \, F_{\lambda}=
\alpha\,\rm{log}\,\lambda+K , 
\end{equation}
 where $K$ is the logarithm of the proportional constant. Figure 5 shows SEDs of TXS 0506+056 at different epochs. From the figure, we confirm that the power law model can represent the SED shape of the blazar well, with the reduced $\chi^{2}$ value of each fit showing the value around one or less. Table 3 lists the fitted parameters of each epoch SED. The spectral slopes $\alpha_{\lambda}$ are plotted as a function of time in Figure 6.

\begin{figure}[ht!]
\epsscale{1.35}
\plotone{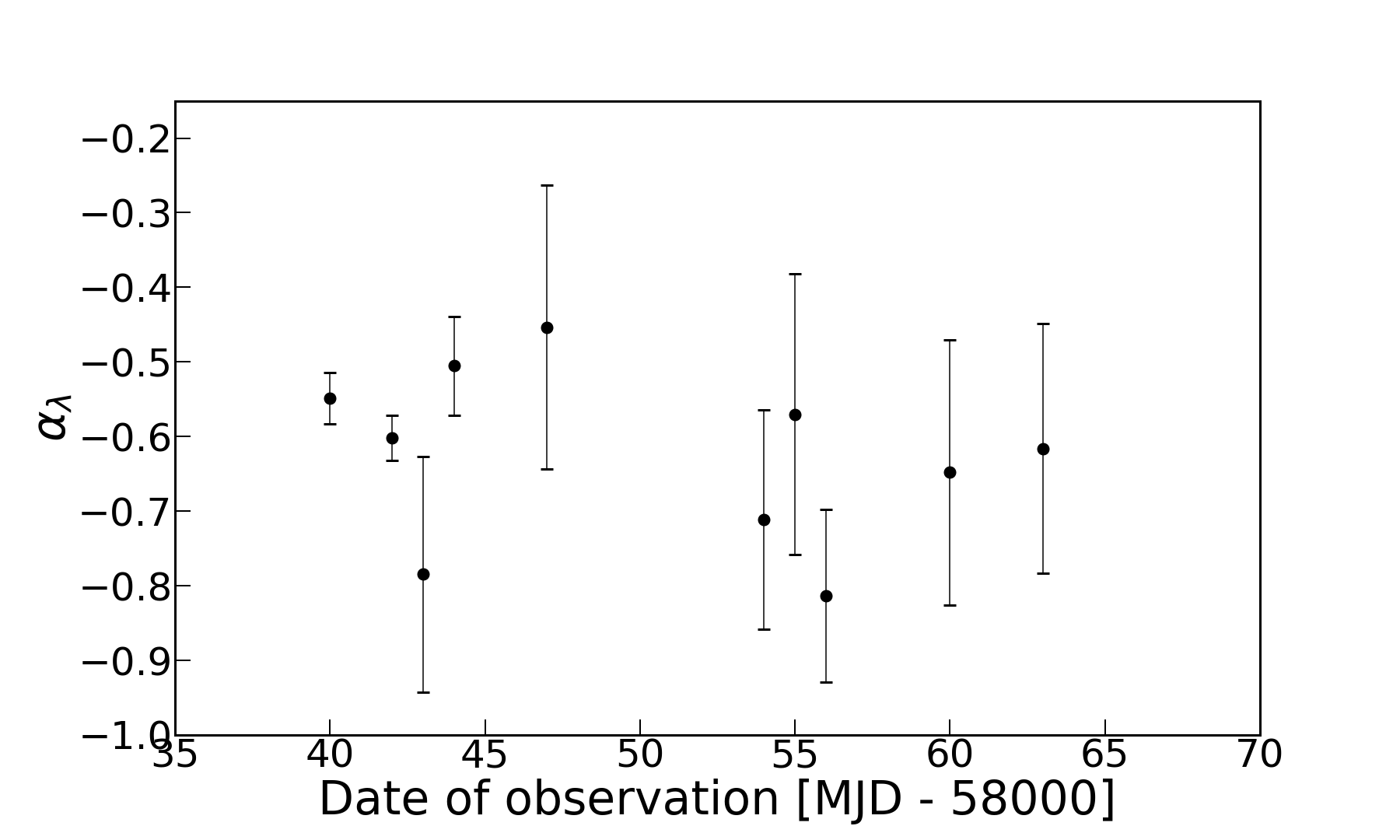}
\caption{The spectral slope as a function of time. The power-law index $\alpha_{\lambda}$ varies during our observation period. \label{fig:fig5}}
\end{figure}
 
 To examine if the power-law index shows a strong variability, we applied the $\chi^2$ test, similarly to what we did in Section 3.1. We find the value of $\chi^2$ = 13.88 with the degree of freedom of 9 for $\alpha_{\lambda}$, and a modest value of $C_{reject}$ = 0.87 for the null hypothesis that $\alpha_{\lambda}$ is not variable.
 This result shows that $\alpha_{\lambda}$, or the spectral shape is not strongly varying above the level that goes beyond the measured uncertainty of $\sim0.15$ during the monitoring period.

 Combining with the result from Section 4.1, we conclude that TXS 0506+056 is variable in short-term ($\sim$ days), but the variability is rather achromatic during the same period.

\section{Discussion}
 
 As we mentioned in the introduction, the blazar nature of TXS 0506+056 is an interesting topic, to better understand the neutrino emission mechanism. We discuss here two spectral properties of TXS 0506+056, its synchrotron peak frequency, $\nu_{peak}^S$, and the spectral variability as a function of brightness.
 
\subsection{Synchrotron peak frequency, $\nu^{S}_{peak}$} 

 The synchrotron peak frequency has been used as a rough way to sort blazars, with FSRQs having log($\nu_{peak}^S$/Hz) $\lesssim$ 14, and BL Lacs having any kind of $\nu_{peak}^S$. Masquerading BL Lacs are known to have log($\nu_{peak}^S$/Hz) $\gtrsim$ 14 (Padovani et al. 2019). It is also a key quantity that defines the blazar sequence. Therefore, we examine its value in detail using our new data.
 
 Due to the moderate but rapid variability of TXS 0506+056, we need to construct its SED with data taken nearly simultaneously ($\lesssim$ 1 day). The $\nu_{peak}^S$ value can be in near-infrared or mid-infrared, therefore it is desirable to have IR data too. Our MB photometry data satisfy the criteria of being taken nearly simultaneously, and we also identify the quasi-simultaneous multi-band photometry data from UV through NIR in Rau et al. (2012, hereafter, R12). For the R12 dataset, the $Swift$ UV and $u, b$, and $v$ and the GROND data ($g,r,i,z,J,H$, and $K$) were taken at about 2 days apart. Therefore, the $Swift$ data and the optical/NIR data are not exactly simultaneous. There is indeed a small offset in the $v$ and $b$ magnitudes versus their GROND $g$ and $r$ magnitudes. Therefore, we shifted the $Swift$ magnitude to match the epoch of the GROND data by deriving the magnitude offset with respect to a reference epoch and applying it to all the $Swift$ data. The achromatic nature of the short-term variability (Section 4.2) justifies this approach for constructing a quasi-simultaneous dataset. The offset was found by deriving the interpolated $v$ magnitude from the GROND data and comparing it to the Swift $v$ mag, where quasi-simultaneous $v$ magnitude is obtained by linearly interpolating the GROND $g$ and $r$ magnitudes and taking the flux at the effective wavelength of the filter. The offset value is found to be 0.14 mag.
 
  Additionally, we estimated the WISE $W1$ (3.36 $\mu$m) and $W2$ (4.61 $\mu$m) data corresponding to the two epochs. Searching for the WISE and NEOWISE data, we find the WISE data taken about 36 and 91 days near our MB observaion and the R12 GROND epochs not close enough in time to treat as quasi-simultaneous. Therefore, we took an alternative approach of finding the correlation between $V$-band data from ASAS-SN, and WISE $W1$ and $W2$ data. We matched the ASAS-SN and WISE data in time domain, allowing a maximum time gap to be 0.2 days. 12 epochs of data are fitted within the criterion. As shown in Figure 7, there is good correlation between $V$ and $W1$ or $W2$, and we derived the relation between the $V$ magnitude and the WISE $W1, W2$ magnitudes as below.
\begin{equation}
W1= 1.30 \, V - 8.63 \,\, (\pm 0.14, 1\,\sigma)
\end{equation}
\begin{equation}
W2= 1.18 \, V - 7.73 \,\, (\pm 0.16, 1\,\sigma)
\end{equation}
Using the above relations, we can calculate the WISE $W1, W2$ magnitude at any epoch using a given $V$ magnitude at a given epoch. The $V$-band derived $W1$ or $W2$ magnitude errors can be obtained as the rms of the offset between the WISE data and the best-fit correlations, which are found to be 0.14 mag and 0.16 mag respectively.
\begin{figure}[ht!]
\epsscale{1.35}
\plotone{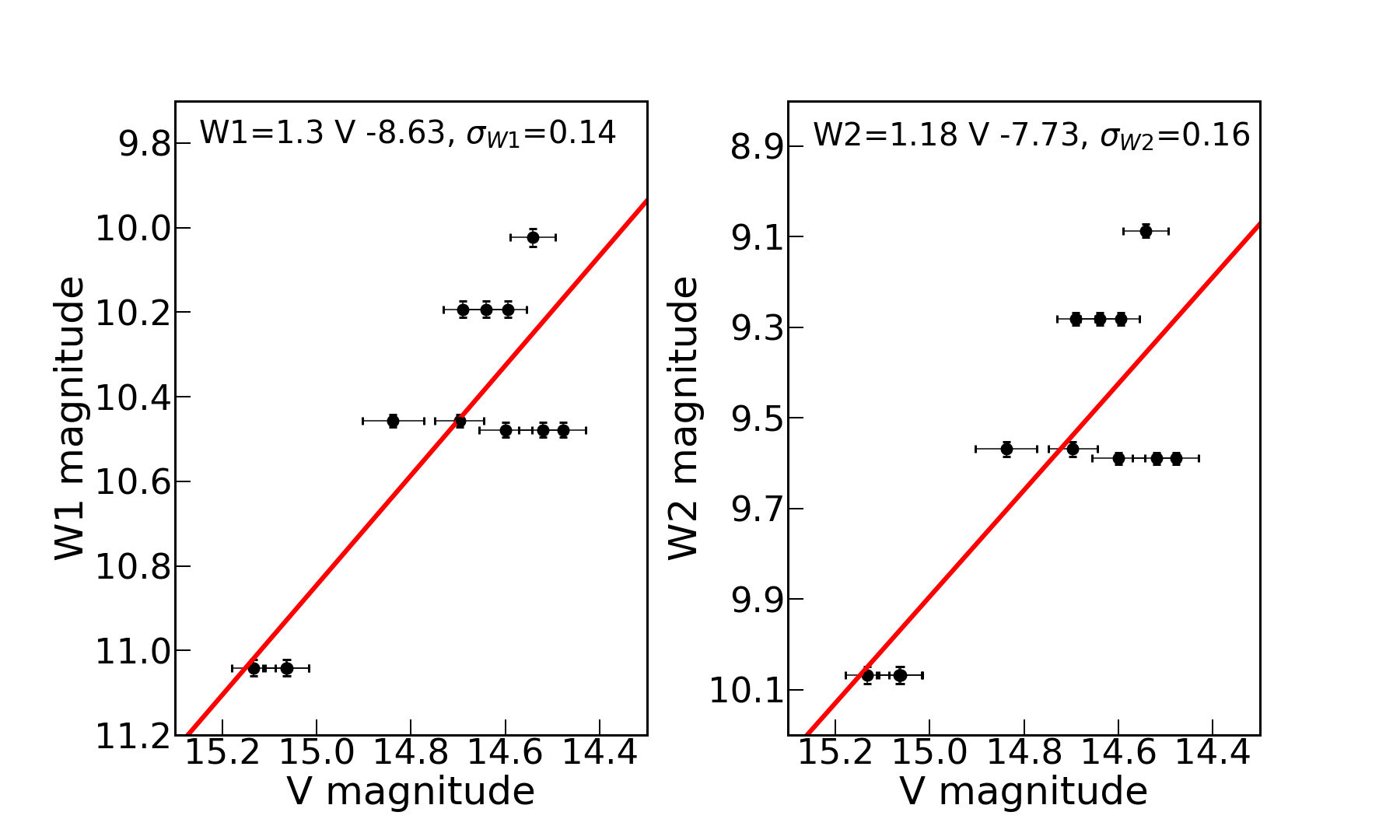}
\caption{The $V$ - $W1$ and $V$ - $W2$ magnitude relation. $V$ magnitudes are from ASAS-SN survey and $W1$, $W2$ magnitudes are from WISE survey. We matched the data collected at 57200 and 58300 MJD with the time difference of observation between WISE and ASAS-SN data less than 0.2 days. The correlation between $V$ vs. $W1$ and $W2$ and the standard deviation in the $WISE$ magnitudes are shown in each plot. \label{fig:fig6}}
\end{figure}

To calculate the WISE $W1$, $W2$ magnitude at the epoch of our observation, we extrapolated the MB SED at the corresponding epoch and took the flux at 550 nm which is the effective central wavelength of the $V$-band filter. For the R12 epoch, the offset-corrected $v$ photometry was adopted to calculate the $W1$ and $W2$ data. In Figure 8, we show these quasi-simultaneous SEDs of the TXS 0506+056. The $V$, $W1$, and $W2$ photometry values are given in Table 4. For our MB data, we used the epoch of 2017 October 14, since the optical SED shape does not vary dramatically during our observation period (Section 4.2). We note that our observation epoch corresponds to the active phase, while the R12 epoch corresponds to a rather inactive phase. This is evident in the long-term $V$-band light curve (Figure 3). The $\nu_{peak}^S$ values were obtained by fitting the SEDs with polynomial functions. We obtained the best-fit $\nu_{peak}^S$ values for the 2$^{nd}$-, 3$^{rd}$-, 4$^{th}$- order polynomial functions, since the derivation of $\nu_{peak}^S$ can vary significantly what fitting function we use. Then, we calculated $\nu_{peak}^S$ and its error to be the mean value and the standard deviation of the three $\nu_{peak}^S$ values obtained from the three fitting methods. We find log($\nu_{peak}^S$)= 14.69 $\pm$ 0.11 for MJD 55540 (the R12 epoch, the darker state) and log($\nu_{peak}^S$) = 14.28 $\pm$ 0.09 for MJD 58040 (our epoch, the brighter state). If we use the magnitudes of Swift and WISE for the darker state that are closest in time (open circles in Figure 8), instead of the quasi-simultaneous magnitudes, we get log($\nu_{peak}^S$) = 14.59 $\pm$ 0.10, which is slightly smaller than the log($\nu_{peak}^S$) from the quasi-simultaneous magnitudes. The $\nu_{peak}^S$ values are consistent with the values presented in P19, and the expected $\nu_{peak}^S$ of a masquerading BL Lac, or an intermediate energy BL Lac. We also note that the $\nu^{S}_{peak}$ determination could have imporved if there were NIR data, since the estimated $\nu^{S}_{peak}$ occurs at NIR for MJD 58040.

\begin{deluxetable}{c|c|c|c}
\tablecaption{Photometry values of $V$, $W1$ and $W2$ filters\label{tab:tab4}}
\tablehead{
\colhead{Date} & \colhead{$V$ (error)} & \colhead{$W1$ (error)} & \colhead{$W2$ (error)}
}
\colnumbers
\startdata
56911.555 & 14.838 (0.065) & 10.457 (0.015) & 9.569 (0.017)\\
56911.556 & 14.697 (0.052) & 10.457 (0.015) & 9.569 (0.017)\\
57276.530 & 15.065 (0.048) & 11.041 (0.019) & 10.068 (0.019)\\
57276.531 & 15.133 (0.046) & 11.041 (0.019) & 10.068 (0.019)\\
57276.532 & 15.062 (0.046) & 11.041 (0.019) & 10.068 (0.019)\\
57441.298 & 14.520 (0.051) & 10.478 (0.018) & 9.589 (0.013)\\
57441.299 & 14.600 (0.056) & 10.478 (0.018) & 9.589 (0.013)\\
57441.301 & 14.478 (0.048) & 10.478 (0.018) & 9.589 (0.013)\\
57805.117 & 14.690 (0.041) & 10.193 (0.019) & 9.282 (0.014)\\
57805.118 & 14.595 (0.040) & 10.193 (0.019) & 9.282 (0.014)\\
57805.119 & 14.640 (0.041) & 10.193 (0.019) & 9.282 (0.014)\\
58004.559 & 14.542 (0.048) & 10.023 (0.021) & 9.087 (0.015)\\
\enddata
\tablecomments{
(1) - Date of observation of $V$-band (Modified Julian Date).;
(2) - $V$-band magnitude from ASAS-SN survey;
(3) - $W1$-band magnitude from WISE survey matched with $V$-band in time within 0.2 days;
(4) - $W2$-band magnitude from WISE survey matched with $V$-band in time within 0.2 days;
}
\end{deluxetable}

\begin{figure}
\epsscale{1.25}
\plotone{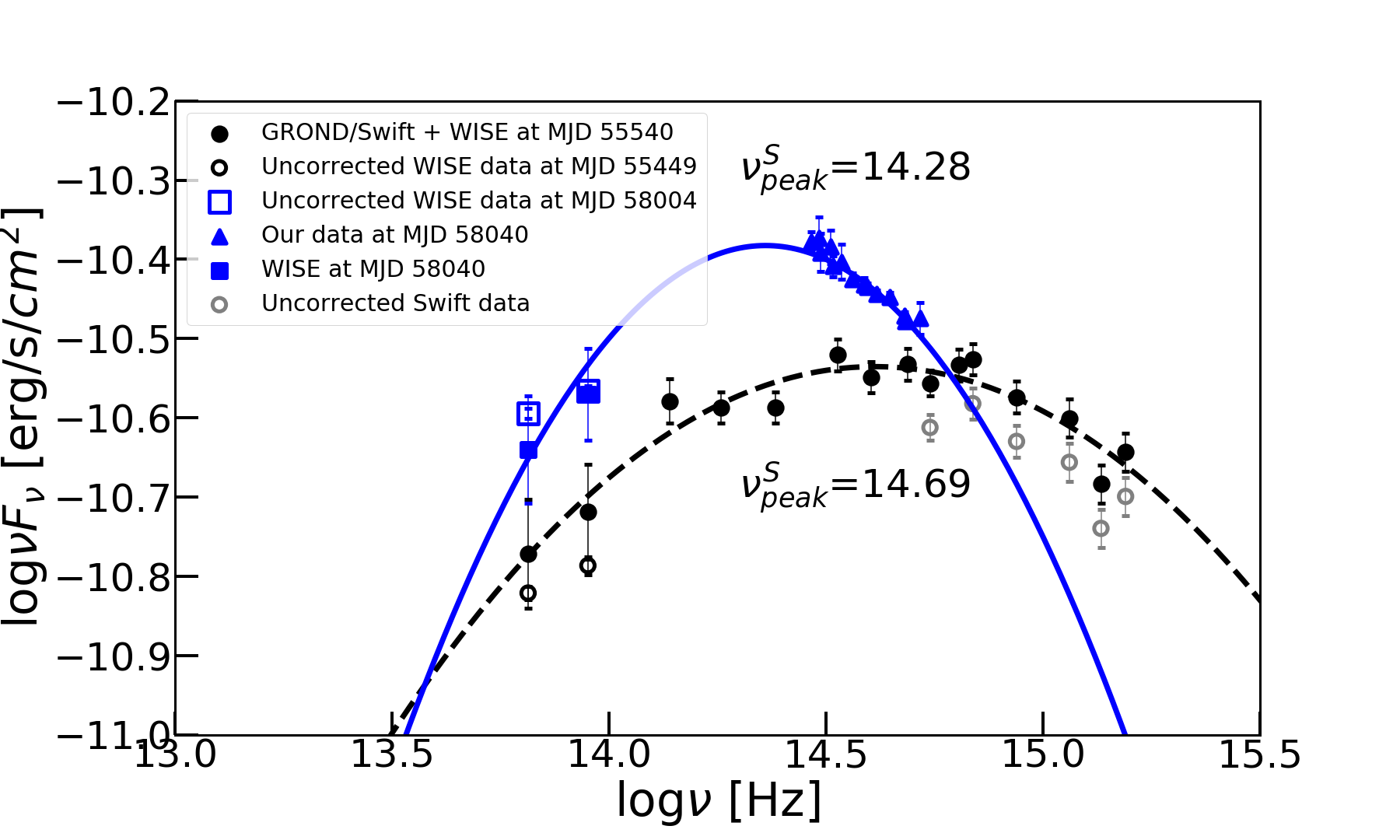}
\caption{SED of the TXS 0506+056 at two different epochs, one near the neutrino emission (brighter phase, the blue line and points), and another at darker phase (the black dashed line and points). The Swift $u, b, v, UVW1, UVM2, UVW2$ fluxes presented in black filled circles are corrected for the 2-days of time gap between GROND and Swift observation. Uncorrected Swift fluxes are also shown in gray empty circles. The WISE data that are closest in time to the epoch of each observation are also shown as open symbols. The best-fit of the data with $2^{nd}$-order polynomial are overploted as solid and dashed lines. \label{fig:fig7}}
\end{figure}

\begin{figure}
\epsscale{1.35}
\plotone{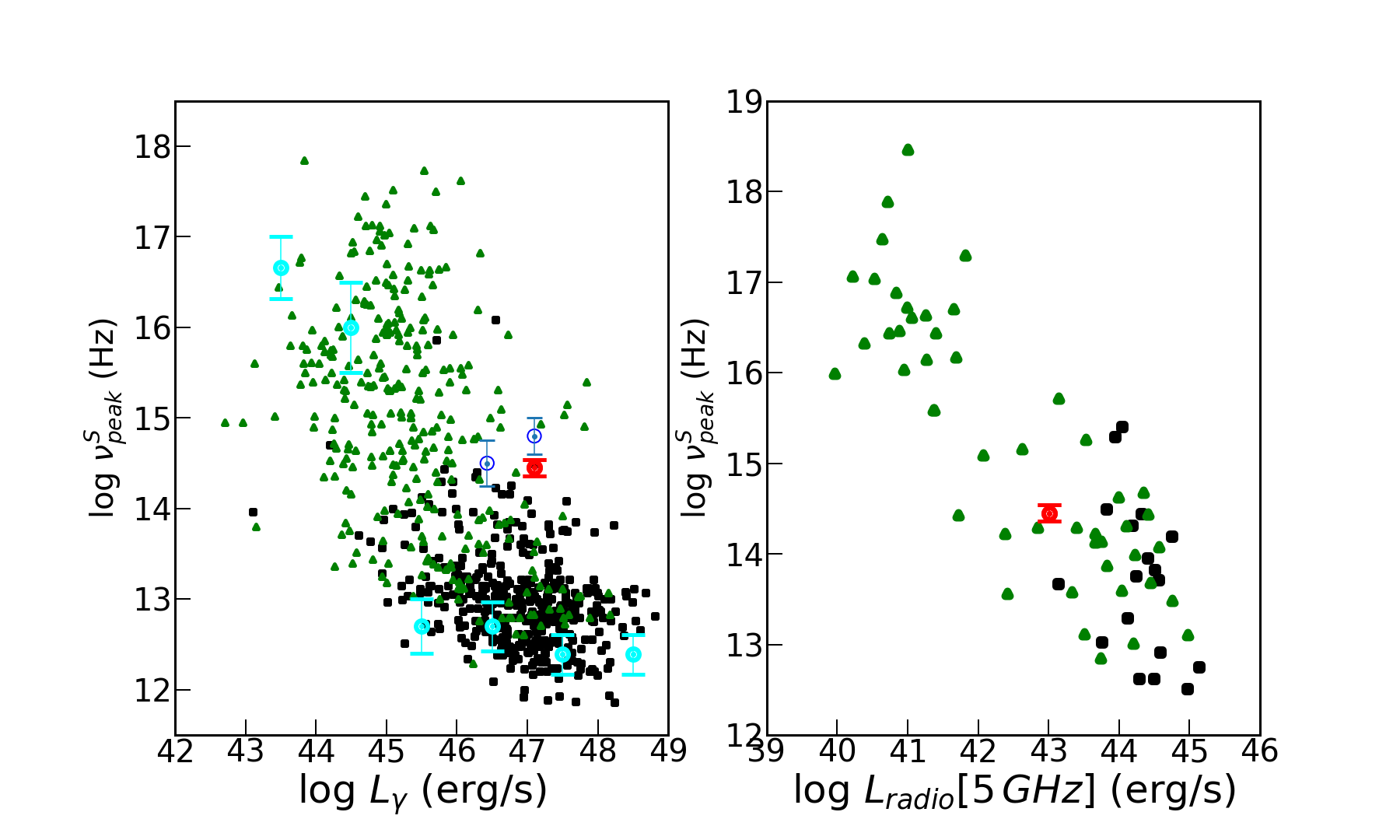}
\caption{The blazar sequence with gamma-ray luminosity (left) and 5 GHz radio luminosity (right). The data of FSRQs and BL Lacs from Fossati et al. (1998, right) and Fermi 3LAC (Ackermann et al. (2015, left) are plotted with black squares and green triangles respectively. The TXS 0506+056 data from P19 is presented with open blue circles and those from our observation is plotted with red circles. The blazar model from Ghisellini et al. (2017) is plotted with cyan circles. \label{fig:fig7}}
\end{figure}
 
\subsection{Is TXS 0506+056 an outlier of the blazar sequence?}
 Figure 9 shows how this object fits in the blazar sequence, in both gamma-ray luminosity vs. $\nu_{peak}^S$ and radio luminosity vs. $\nu_{peak}^S$ spaces. For the case of the gamma-ray luminosity vs. $\nu_{peak}^S$ plot, the FSRQ and BL Lac data are taken from the Fermi 3LAC catalog (Ackermann et al. 2015), while the Fossati et al. (1998) data are used for the radio luminosity vs. $\nu_{peak}^S$ plot. The gamma-ray luminosity of TXS 0506+056 is taken from P19. The 5 GHz radio flux of TXS 0506+056 is derived from the interpolation of 4.68 GHz and 7.02 GHz data (rest-frame) of IceCube Collaboration et al. (2018) obtained at MJD = 58038.35, which is the closest in time to our MB observation. The luminosity ($\rm{log}\,L_{5GHz}$) is calculated from flux, assuming a $\Lambda$CDM concordance cosmology of $H_{0} = 70$ km s$^{-1}$ Mpc$^{-1}$, $\Omega_{M} = 0.3$ and $\Omega_{\Lambda} = 0.7$ which has been supported by observations in the past decades (e.g., Im et al. 1997; Planck Collaboration et al. 2018). We found $\rm{log}\,L_{5GHz}$ = 43.01. On contrary to the suggestion by P19, Figure 9 shows that TXS 0506+056 sits in the blazar sequence, meaning that TXS 0506+056 is not an exceptional blazar in the blazar sequence. Compared to the blazar sequence constructed from the blazar model of Ghisellini et al. (2017, cyan circles), TXS 0506+056 appears as an outlier, but TXS 0506+056 sits near the Ackermann et al. (2015, black squares and green triangles) data, which is also used to construct the model of Ghisellini et al. (2017). We note that the blazar sequence points taken from Ghisellini et al. (2017) deviates from the Fermi 3LAC SEDs and the Fossati et al. (1998) points. By plotting the Fermi 3LAC SEDs of blazars, we confirm that the values listed in the Fermi 3LAC catalogs. Therefore, it is not clear the origin of the discrepancy, but we suspect that it is due to the model function used to determine the $\nu_{peak}^S$ by Ghisellini et al. (2017).

\subsection{Redder-when-brighter or bluer-when-brighter?}
 The next interesting point in Figure 8 is the difference between the spectral shapes during the active and inactive phases. We find that the R12 spectral shape in the inactive phase is flatter in the optical than the active phase (our data), suggesting a redder-when-brighter behavior found in FSRQs. The redder-when-brighter behavior is likely to be due to the thermal component of the SED at optical-IR range as suggested by many previous studies (Rani et al. 2010; Bonning et al. 2012). The SED of the blazars at optical-IR range can be broken into two main components. One is the non-thermal emission by the jet and the other is the thermal emission by the accretion disk. Intrinsically bright quasars such as FSRQs, have a strong thermal emission component with a blue SED, and their rapid variability is controlled by a non-thermal component which is dominant at longer wavelength. Therefore, when the jet activity is high, the non-thermal component brightens (SED at longer wavelength) causing the redder-when-brighter behavior. Therefore, the redder-when-brigther behavior of TXS 0506+056 suggests that it is an intrinsically bright AGN. Obviously, one should not draw a definitive conclusion from the comparison of SEDs at only the two epochs. Therefore, we examined how the optical spectral slope changes with the apparent magnitude, assuming that the apparent magnitude is a good proxy of the AGN activity. Figure 10 shows $\alpha_{\lambda}$ versus $m575$ and $m775$ magnitudes which are the bluest and reddest passbands covered by WIT. When SQUEAN MB photometry is available we use it instead of the WIT photometry since it provides smaller errors. The Pearson's coefficient ($\rho$) and its p-value were 0.38 and 0.35 for the case of $m575$-band and -0.269 and 0.45 for the case of $m775$-band. Therefore, the significance of the correlation is only 65\% and 55\% each. The results shown in Figure 10 indicate that the change in $\alpha_{\lambda}$ is consistent with null hypothesis within the measurement error, agreeing with the achromatic behavior of the SED shape over a short period found in Section 4.2. It appears that a longer time and larger magnitude base line is desirable to better characterize the spectral variability. If there was such a long-term data, since the measurement accuracy for the $\alpha_{\lambda}$ of MB SED is $\delta \alpha_{\lambda}$ of 0.05 - 0.2 (Table 3), the SED slope change of $\alpha_{\lambda} \sim -0.6$ (our result) to $\alpha_{\lambda} \sim -1.0$ (R12) would have been easily explored.

\subsection{Implications on the neutrino emission mechanism}
 TXS 0506+056 being an intrinsically bright AGN is in line with the conventional view of high energy neutrino production in AGN jets (e.g., see discussion in Ansoldi et al. 2018). The typical production channel of high energy neutrinos is the decay of charged pions that were produced in collisions between high energy protons in jets and ambient matters (Halzen 2017; Meszaros 2017), such as matter ($pp$ interaction) or low energy photons ($p\gamma$ interaction). In AGN jets, with their low-density plasma, the $p\gamma$ interaction has been a more favored channel (Mannheim 1995) where the target photon energies are in the range of UV to soft X-ray photons. Hot, luminous accretion disks of intrinsically bright AGNs are an excellent source for the UV to soft X-ray photons (Ansoldi et al. 2018). Hence, FSRQ-type blazars have been favored to be high energy neutrino sources (Mannheim et al. 1992; Atoyan \& Dermer 2001; Murase et al. 2014; Dermer et al. 2014), and our observation is consistent with such an expectation.

\begin{figure}
\epsscale{1.3}
\plotone{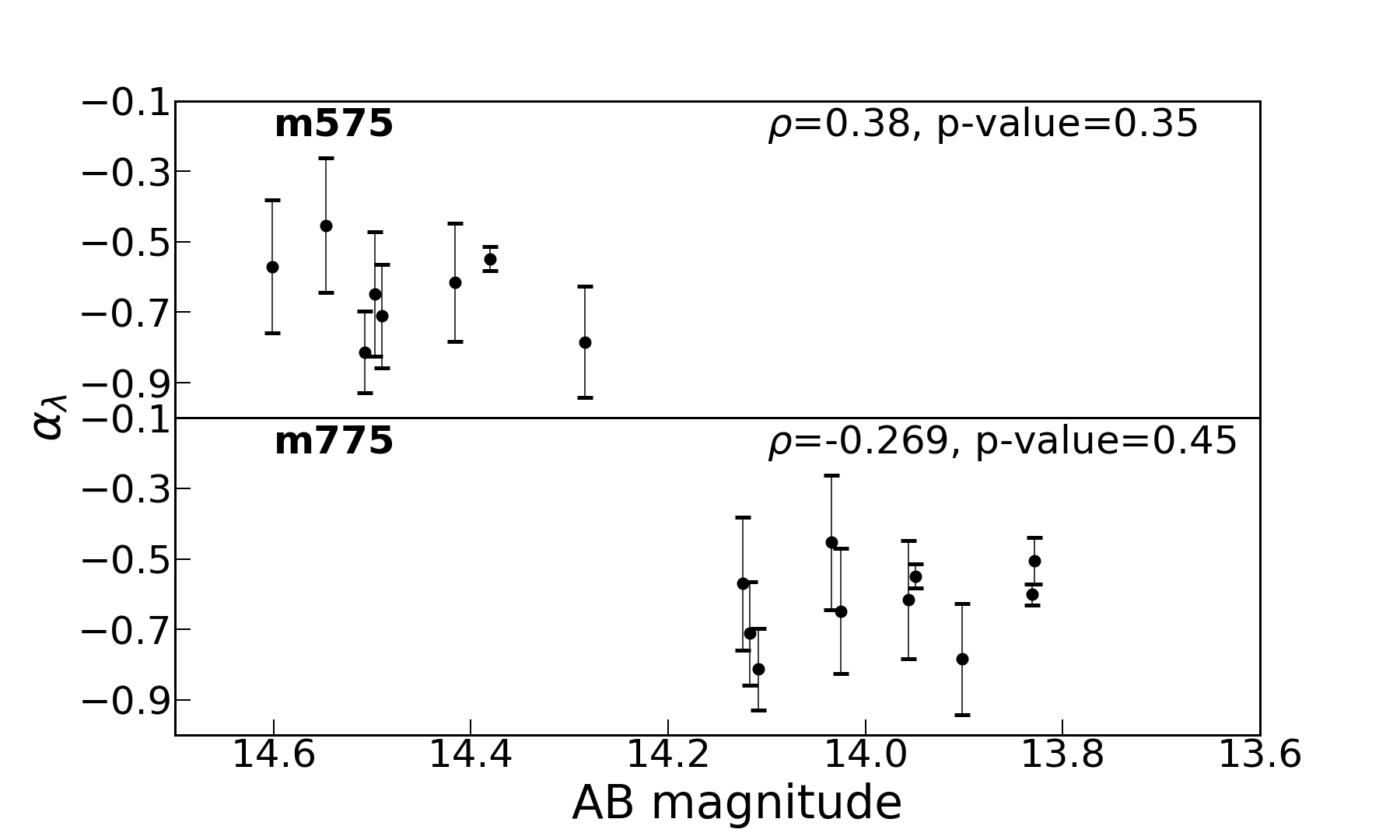}
\caption{Upper pannel shows the $\alpha_{\lambda}$ versus $m575$ band magnitude plot and lower pannel shows the $\alpha_{\lambda}$ versus $m775$ band magnitude plot. The Pearson's coefficient ($\rho$) and its p-value are shown as well. \label{fig:fig8}}
\end{figure}

\subsection{Discussion summary}
 The above analyses teach us that TXS 0506+056 is not an extraordinary outlier from the blazar sequence. From the long-term spectral variability analysis, we find that TXS 0506+056 varies as redder-when-brighter which is consistent with the earlier P19 report that TXS 0506+056 is not a BL Lac. However, this was limited to only two epochs. A more definitive answer requires data at multiple long-term epochs. Figure 10 and the spectral slope measurement accuracy achieved from our observation gives a great promise that spectral variability study with MB data at more extended epochs could firmly establish if the object is redder-when-brighter or bluer-when-brighter.

\section{Conclusion}

 We conducted the time-series, MB observational study of the blazar, TXS 0506+056, to understand its nature as the host galaxy of the neutrino event IceCube-170922A. Our observation reveals the variable nature of TXS 0506+056, and the MB data obtained using WIT and SQUEAN are found to excellently trace the optical SED shape of the blazar allowing us to measure the spectral index, $\alpha_{\lambda}$ to be $\sim -0.6$ to $\sim -0.8$ to the accuracy of 2 \% to 10 \%. Our MB data were combined with published dataset and archival data to produce quasi-simultaneous SED at two epochs (one active, another inactive). The multi-wavelength SEDs show that $\nu_{peak}^S$ is of order of $\nu_{peak}^S \sim 10^{14.5}$ Hz, which is consistent with previous results. Considering that the $\nu_{peak}^{S}$ value is in NIR, NIR monitoring observation can help better determine $\nu_{peak}^{S}$. However, the comparison with the blazar sequence data in both gamma-ray and radio luminosities shows that TXS 0506+056 is not an outlier of the blazar sequence. The comparison of the 2012 data (inactive) and the 2017 data (active) shows a flatter SED at 2012, possibly indicating the redder-when-brighter nature of FSRQ for TXS 0506+056, but the spectral variability analysis using our data show no clear spectral change as a function of the brightness. A more extensive study using a larger set of data is desirable, using a longer time-baseline and larger phase difference.
 
 In conclusion, our study demonstrates that the MB photometry observation with small and medium-sized telescopes can be a powerful way to study spectral variability of many kinds of astrophysical objects such as neutrino sources, AGNs, gravitational-wave sources, gamma-ray bursts, and supernovae, to name a few (e.g., Abbott et al. 2017; Im et al. 2015, 2019; Kim et al. 2019a). Additional NIR monitoring data will be also helpful for better determining the spectral shape, especially $\nu_{peak}^{S}$ of blazars like TXS 0506+056. Wide-field telescopes with MB filters, such as WIT with a field of view of 5.5 deg$^2$, can be a great force for multi-messenger astronomy to search for and characterize optical counterparts and host galaxies of neutrino and gravitaional-wave sources (e.g., Abbott et al. 2017; Im et al. 2017) for which initial positional estimates can be very uncertain ($\gg 1$ deg$^2$).

\acknowledgements
 We thank John Kuehne and the staffs at the McDonald Observatory for their support during the installation and the operation of WIT. This work was supported by the National Research Foundation of Korea (NRF) grant (2020R1A2C3011091) by the Korean government (MSIT) and the Korea Astronomy and Space Science Institute under the R\&D program (Project No. 2020-1-600-05) supervised by MSIT. S.S. acknowledges support from Basic Science Research Program through the National Research Foundation of Korea (NRF) funded by the Ministry of Education (No.2020R1A6A3A13069198). This paper includes data taken at the McDonald Observatory of the University of Texas at Austin.

\facilities{Struve (SQUEAN)}
\software{Astronomy.net (Lang et al. 2010),\ IRAF (Tody 1986),\ SExtractor (Bertin \& Arnouts 1996),\ SWarp (Bertin 2010)}

\appendix
\section{Photometry Information Tables}
\begin{deluxetable*}{c|c|c|c|c|c|c|c|c}
\tablecaption{TXS 0506+056 photometry with WIT\label{tab:tab2}}
\tablehead{
\colhead{WIT}\\
\colhead{Date} & \colhead{Weather} & \colhead{Filter} & \colhead{Seeing} & \colhead{Exposure time} & \colhead{Zeropoint (error)} & \colhead{Detection limit} & \colhead{mag} & \colhead{$\sigma_{m}$}}
\colnumbers
\startdata
2017-10-14 08:31 & clear & $m575$ & 4.06 & 150s$\times$4 & 18.409 (0.045) & 17.58 & 14.38 & 0.05\\
2017-10-14 08:51 & clear & $m625$ & 4.39 & 150s$\times$4 & 18.212 (0.036) & 17.37 & 14.35 & 0.04\\
2017-10-14 09:13 & clear & $m675$ & 4.84 & 150s$\times$4 & 17.800 (0.048) & 16.90 & 14.17 & 0.05\\
2017-10-14 09:25 & clear & $m725$ & 4.55 & 150s$\times$4 & 17.306 (0.041) & 16.58 & 14.07 & 0.04\\
2017-10-14 09:40 & clear & $m775$ & 4.95 & 150s$\times$4 & 16.759 (0.046) & 16.29 & 14.01 & 0.05\\
2017-10-17 07:35 & clear & $m575$ & 5.18 & 150s$\times$4 & 18.412 (0.042) & 17.34 & 14.28 & 0.04\\
2017-10-17 07:46 & clear & $m625$ & 5.09 & 150s$\times$4 & 18.195 (0.033) & 17.34 & 14.16 & 0.03\\
2017-10-17 07:56 & clear & $m675$ & 4.95 & 150s$\times$4 & 17.8073 (0.046) & 17.17 & 14.04 & 0.05\\
2017-10-17 08:07 & clear & $m725$ & 5.86 & 150s$\times$4 & 17.386 (0.032) & 16.66 & 13.94 & 0.04\\
2017-10-17 08:19 & clear & $m775$ & 5.64 & 150s$\times$4 & 16.768 (0.033) & 16.18 & 13.90 & 0.04\\
2017-10-21 09:47 & clear & $m575$ & 3.70 & 150s$\times$4 & 18.404 (0.050) & 17.78 & 14.55 & 0.05\\
2017-10-21 09:58 & clear & $m625$ & 4.12 & 150s$\times$4 & 18.177 (0.034) & 17.58 & 14.40 & 0.04\\
2017-10-21 10:10 & clear & $m675$ & 4.46 & 150s$\times$4 & 17.786 (0.049) & 17.19 & 14.26 & 0.05\\
2017-10-21 10:21 & clear & $m725$ & 3.99 & 150s$\times$4 & 17.312 (0.042) & 16.95 & 14.17 & 0.05\\
2017-10-21 10:33 & clear & $m775$ & 4.17 & 150s$\times$4 & 16.685 (0.045) & 16.39 & 14.03 & 0.05\\
2017-10-28 08:27 & clear & $m575$ & 3.62 & 150s$\times$5 & 18.388 (0.044) & 14.81 & 14.49 & 0.05\\
2017-10-28 08:54 & clear & $m625$ & 3.94 & 150s$\times$5 & 18.099 (0.026) & 15.20 & 14.46 & 0.03\\
2017-10-28 09:09 & clear & $m675$ & 4.20 & 150s$\times$5 & 17.770 (0.036) & 14.83 & 14.31 & 0.04\\
2017-10-28 09:24 & clear & $m725$ & 3.86 & 150s$\times$5 & 17.280 (0.029) & 14.44 & 14.22 & 0.03\\
2017-10-28 09:40 & clear & $m775$ & 4.14 & 150s$\times$5 & 16.688 (0.027) & 13.84 & 14.12 & 0.05\\
2017-10-29 07:49 & clear & $m575$ & 4.01 & 150s$\times$5 & 18.426 (0.045) & 17.91 & 14.60 & 0.04\\
2017-10-29 08:05 & clear & $m625$ & 4.38 & 150s$\times$5 & 18.203 (0.038) & 17.60 & 14.47 & 0.04\\
2017-10-29 08:20 & clear & $m675$ & 4.55 & 150s$\times$5 & 17.798 (0.051) & 17.34 & 14.29 & 0.05\\
2017-10-29 08:36 & clear & $m725$ & 4.25 & 150s$\times$5 & 17.360 (0.044) & 17.16 & 14.26 & 0.05\\
2017-10-29 08:51 & clear & $m775$ & 4.68 & 150s$\times$5 & 16.768 (0.049) & 16.62 & 14.12 & 0.05\\
2017-10-30 06:17 & clear & $m575$ & 6.05 & 150s$\times$4 & 18.298 (0.033) & 16.81 & 14.51 & 0.04\\
2017-10-30 06:30 & clear & $m625$ & 5.89 & 150s$\times$4 & 18.097 (0.020) & 16.91 & 14.37 & 0.02\\
2017-10-30 06:42 & clear & $m675$ & 5.22 & 150s$\times$4 & 17.686 (0.036) & 16.83 & 14.26 & 0.04\\
2017-10-30 06:53 & clear & $m725$ & 7.63 & 150s$\times$4 & 17.440 (0.009) & 16.30 & 14.19 & 0.02\\
2017-10-30 07:12 & clear & $m775$ & 6.53 & 150s$\times$4 & 16.756 (0.012) & 15.97 & 14.11 & 0.03\\
2017-11-03 05:58 & clear & $m575$ & 4.37 & 150s$\times$4 & 18.440 (0.045) & 16.27 & 14.50 & 0.05\\
2017-11-03 06:10 & clear & $m625$ & 4.69 & 150s$\times$4 & 18.201 (0.030) & 16.28 & 14.36 & 0.03\\
2017-11-03 06:22 & clear & $m675$ & 4.51 & 150s$\times$4 & 17.829 (0.044) & 16.35 & 14.22 & 0.05\\
2017-11-03 06:33 & clear & $m725$ & 4.65 & 150s$\times$4 & 17.376 (0.030) & 16.23 & 14.18 & 0.04\\
2017-11-03 06:45 & clear & $m775$ & 4.87 & 150s$\times$4 & 16.777 (0.047) & 15.92 & 14.02 & 0.05\\
2017-11-06 06:01 & clear & $m575$ & 6.64 & 150s$\times$4 & 18.488 (0.035) & 15.86 & 14.42 & 0.04\\
2017-11-06 06:16 & clear & $m625$ & 5.37 & 150s$\times$5 & 18.195 (0.026) & 16.16 & 14.31 & 0.03\\
2017-11-06 06:39 & clear & $m675$ & 5.80 & 150s$\times$5 & 17.850 (0.040) & 15.97 & 14.14 & 0.04\\
2017-11-06 07:00 & clear & $m725$ & 4.87 & 150s$\times$4 & 17.369 (0.039) & 15.93 & 14.11 & 0.04\\
2017-11-06 07:22 & clear & $m775$ & 5.07 & 150s$\times$5 & 16.774 (0.047) & 15.60 & 13.96 & 0.05\\
\enddata
\tablecomments{
(1) - Date of observation (UT);
(2) - Weather;
(3) - Name of filter;
(4) - Seeing value ($\arcsec$);
(5) - Exposure time (sec);
(6) - Zeropoint (mag);
(7) - Point source detection limit at 5-$\sigma$ with exposure time and zeropoint in same row (mag);
(8) - Magnitude;
(9) - Error of the magnitude
}
\end{deluxetable*}
\begin{deluxetable*}{c|c|c|c|c|c|c|c|c}
\tablecaption{TXS 0506+056 photometry with 2.1m telescope\label{tab:tab3}}
\tablehead{
\colhead{2.1m telescope}\\
\colhead{Date} & \colhead{Weather} & \colhead{Filter} & \colhead{Seeing} & \colhead{Exposure time} & \colhead{Zeropoint (error)} & \colhead{Detection limit} & \colhead{mag} & \colhead{$\sigma_{m}$}
}
\colnumbers
\startdata
2017-10-14 09:07 & clear & $Y$ & 1.73 & 60s$\times$3 & 22.414 (0.067) & 19.55 & 13.54 & 0.07\\
2017-10-14 09:22 & clear & $i$ & 2.53 & 10s$\times$36 & 24.979 (0.012) & 20.72 & 13.97 & 0.01\\
2017-10-14 09:02 & clear & $m1025$ & 1.86 & 60s$\times$3 & 21.046 (0.032) & 18.61 & 13.51 & 0.03\\
2017-10-14 08:29 & clear & $m625$ & 1.49 & 30s$\times$5 & 23.087 (0.012) & 20.52 & 14.28 & 0.01\\
2017-10-14 08:32 & clear & $m675$ & 1.47 & 30s$\times$5 & 23.436 (0.015) & 20.74 & 14.14 & 0.02\\
2017-10-14 08:36 & clear & $m725$ & 1.51 & 15s$\times$12 & 23.675 (0.012) & 20.64 & 14.05 & 0.01\\
2017-10-14 08:40 & clear & $m775$ & 1.66 & 15s$\times$12 & 23.670 (0.021) & 20.49 & 13.95 & 0.02\\
2017-10-14 08:45 & clear & $m825$ & 1.63 & 15s$\times$12 & 23.642 (0.019) & 20.43 & 13.86 & 0.02\\
2017-10-14 08:49 & clear & $m875$ & 1.77 & 30s$\times$6 & 23.459 (0.056) & 20.44 & 13.75 & 0.06\\
2017-10-14 08:54 & clear & $m925s$ & 1.48 & 30s$\times$6 & 22.375 (0.053) & 19.30 & 13.64 & 0.05\\
2017-10-14 08:58 & clear & $m975$ & 1.66 & 60s$\times$3 & 22.113 (0.060) & 19.65 & 13.60 & 0.06\\
2017-10-14 09:29 & clear & $r$ & 2.15 & 15s$\times$12 & 24.318 (0.008) & 20.80 & 14.30 & 0.01\\
2017-10-14 09:12 & clear & $z$ & 1.31 & 15s$\times$12 & 23.708 (0.033) & 20.38 & 13.71 & 0.03\\
2017-10-16 08:54 & clear & $Y$ & 1.30 & 30s$\times$10 & 21.849 (0.046) & 19.29 & 13.42 & 0.05\\
2017-10-16 09:08 & clear & $i$ & 2.46 & 10s$\times$30 & 24.908 (0.047) & 20.47 & 13.92 & 0.05\\
2017-10-16 08:45 & clear & $m1025$ & 1.52 & 60s$\times$5 & 20.600 (0.020) & 18.55 & 13.42 & 0.02\\
2017-10-16 07:51 & clear & $m625$ & 2.52 & 60s$\times$5 & 23.056 (0.009) & 20.51 & 14.16 & 0.01\\
2017-10-16 07:58 & clear & $m675$ & 1.89 & 30s$\times$10 & 23.311 (0.010) & 20.69 & 14.02 & 0.01\\
2017-10-16 08:06 & clear & $m725$ & 1.71 & 15s$\times$20 & 23.595 (0.009) & 20.65 & 13.95 & 0.01\\
2017-10-16 08:12 & clear & $m775$ & 1.61 & 15s$\times$20 & 23.538 (0.035) & 20.51 & 13.83 & 0.04\\
2017-10-16 08:19 & clear & $m825$ & 1.58 & 15s$\times$20 & 23.373 (0.054) & 20.40 & 13.70 & 0.05\\
2017-10-16 08:28 & clear & $m875$ & 1.59 & 15s$\times$20 & 23.165 (0.046) & 20.11 & 13.61 & 0.05\\
2017-10-16 08:33 & clear & $m925s$ & 1.86 & 60s$\times$5 & 22.494 (0.044) & 19.38 & 13.52 & 0.04\\
2017-10-16 08:39 & clear & $m975$ & 1.44 & 30s$\times$10 & 21.802 (0.026) & 19.44 & 13.48 & 0.03\\
2017-10-16 09:15 & clear & $r$ & 1.84 & 10s$\times$30 & 23.994 (0.030) & 20.55 & 14.13 & 0.03\\
2017-10-16 09:00 & clear & $z$ & 1.30 & 10s$\times$30 & 23.327 (0.026) & 19.92 & 13.59 & 0.03\\
2017-10-18 11:53 & cloudy & $Y$ & 1.13 & 60s$\times$8 & 21.475 (0.059) & 19.27 & 13.41 & 0.06\\
2017-10-18 12:07 & cloudy & $i$ & 1.23 & 10s$\times$18 & 23.336 (0.034) & 18.72 & 13.85 & 0.03\\
2017-10-18 11:43 & cloudy & $m1025$ & 1.05 & 60s$\times$10 & 18.749 (0.089) & 17.34 & 13.39 & 0.09\\
2017-10-18 10:24 & cloudy & $m625$ & 1.27 & 60s$\times$5 & 19.307 (0.028) & 17.81 & 14.16 & 0.03\\
2017-10-18 10:31 & cloudy & $m675$ & 1.18 & 60s$\times$12 & 20.156 (0.038) & 18.89 & 14.05 & 0.04\\
2017-10-18 10:42 & cloudy & $m725$ & 1.09 & 60s$\times$5 & 19.284 (0.018) & 17.68 & 13.91 & 0.02\\
2017-10-18 10:53 & cloudy & $m775$ & 1.11 & 60s$\times$8 & 19.306 (0.028) & 17.72 & 13.83 & 0.03\\
2017-10-18 11:02 & cloudy & $m825$ & 1.46 & 60s$\times$10 & 18.948 (0.049) & 17.01 & 13.76 & 0.05\\
2017-10-18 11:15 & cloudy & $m875$ & 1.04 & 60s$\times$10 & 22.212 (0.069) & 20.03 & 13.58 & 0.07\\
2017-10-18 11:23 & cloudy & $m925s$ & 1.84 & 60s$\times$5 & 21.462 (0.037) & 18.75 & 13.52 & 0.04\\
2017-10-18 11:31 & cloudy & $m975$ & 1.03 & 60s$\times$10 & 20.042 (0.094) & 18.47 & 13.46 & 0.09\\
2017-10-18 12:11 & cloudy & $r$ & 1.87 & 30s$\times$10 & 22.270 (0.031) & 18.33 & 14.19 & 0.03\\
2017-10-18 12:01 & cloudy & $z$ & 1.30 & 15s$\times$12 & 22.833 (0.032) & 19.16 & 13.57 & 0.03\\
\enddata

\tablecomments{
(1) - Date of observation (UT);
(2) - Weather;
(3) - Name of filter;
(4) - Seeing value ($\arcsec$);
(5) - Exposure time (sec);
(6) - Zeropoint (mag);
(7) - Point source detection limit at 5-$\sigma$ with exposure time and zeropoint in same row (mag);
(8) - Magnitude;
(9) - Error of the magnitude
}

\end{deluxetable*}

\end{document}